\documentclass{article}
\usepackage{spconf,amsmath,amsfonts,mathtools,graphicx,hyperref,xcolor,algorithm,algpseudocode}

\DeclareMathOperator*{\argmin}{argmin}

\newcommand{\data}{N}
\newcommand{\idx}{n}
\newcommand{\len}{L}
\newcommand{\comp}{\ell}
\newcommand{\unif}{\text{Uniform}}
\newcommand{\truesig}{X_{\hspace{-0.2mm}*}}

\newcommand{\Fig}{Fig.}

\title{Moment-based Posterior Sampling for Multi-reference Alignment}

\name{Axel Janson, Joakim Andén\thanks{This work was funded by the Swedish Research council through grant no. 2023-04143 and the computations were enabled by resources provided by the National Academic Infrastructure for Supercomputing in Sweden (NAISS), partially funded by the Swedish Research Council through grant agreement no. 2022-06725.}}
\address{KTH Royal Institute of Technology\\Department of Mathematics}
\begin{document}
%
\maketitle
\begin{abstract}
We propose a Bayesian approach to the problem of multi-reference alignment -- the recovery of signals from noisy, randomly shifted observations. While existing frequentist methods accurately recover the signal at arbitrarily low signal-to-noise ratios, they require a large number of samples to do so. In contrast, our proposed method leverages diffusion models as data-driven plug-and-play priors, conditioning these on the sample power spectrum (a shift-invariant statistic) enabling both accurate posterior sampling and uncertainty quantification. The use of an appropriate prior significantly reduces the required number of samples, as illustrated in simulation experiments with comparisons to state-of-the-art methods such as expectation--maximization and bispectrum inversion. These findings establish our approach as a promising framework for other orbit recovery problems, such as cryogenic electron microscopy (cryo-EM).
\end{abstract}
\section{Introduction} \label{sec:introduction}

Multi-reference alignment (MRA) is the problem of recovering a signal~$X$ from a dataset of~$\data$ observations, subject to random periodic shifts and additive noise. The signal is represented as samples on a uniform grid of size~$\len,$ viewed as a vector~$X \in \mathbb{R}^\len$ which is then shifted by the cyclic group of order~$\len.$ Each observation~$Y_\idx$ is then given by
\begin{equation} \label{eq:problem_MRA}
    Y_\idx = S^{\,\Phi_{\!\idx}} \! X + \varepsilon_\idx, \quad \idx \in \{ 1, \ldots, \data \},
\end{equation}
where~$S$ is the shift-by-one matrix,~$\varepsilon_\idx \sim \mathcal{N}(0, \sigma^2 I)$,~$\sigma^2 \geq 0$, and~$\Phi_\idx \sim \unif(\{0, ..., \len-1\}).$ Note that in this formulation,~$X$ can only be recovered up to global shift, i.e., we only recover the \emph{orbit} of~$X.$ MRA can therefore be seen as an \emph{orbit recovery problem}.

\begin{figure}[t]
\begin{minipage}[t]{\linewidth}
  \centering
  \centerline{\includegraphics[trim=10 100 10 100, clip, width=8cm]{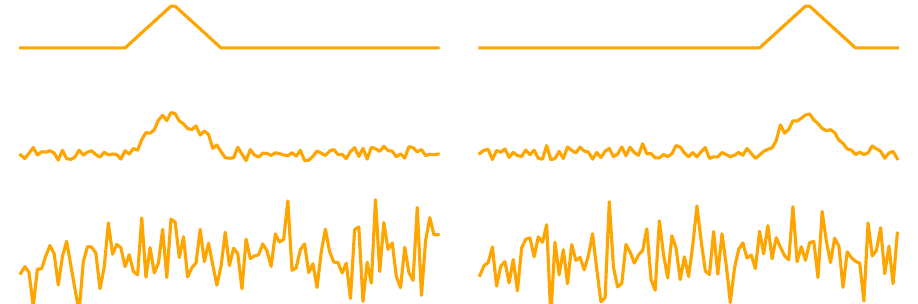}}
\end{minipage}
\caption{Illustrative MRA observations given by~\eqref{eq:problem_MRA}, with the same underlying signal, the same shift across each column, and increasing variance~$\sigma^2$ along rows.}
\label{fig:mra_example}
\end{figure}

The main obstacle that orbit recovery methods for MRA must overcome is that additive noise interferes with shift estimation. For example, if all of the shifts~$\Phi_\idx$ were known, orbit recovery could be performed with a simple align-and-average approach, with sample complexity scaling as~$\sigma^2.$ However, with unknown shifts and generic~$X,$ the sample complexity of any estimator rises to~$\sigma^6$ for small signal-to-noise ratios, i.e.,~$\sigma^2 \gg \|X\|^2$~\cite{samplecomplexity_mra}. 
To better understand how additive noise complicates orbit recovery, see \Fig~\ref{fig:mra_example}, which illustrates how shift estimation can be reliable in the low-noise regime, yet almost completely random in the high-noise regime.

MRA exemplifies the fundamental statistical estimation problem that arises when attempting to recover a signal from a set of noisy observations, corrupted by the hidden action of a known group. This general description captures challenges common to a number of scientific and engineering domains, including robotics and structural biology~\cite{SLAMrobotics, medicinemicroscopy}. In particular, MRA shares many properties with single-particle cryogenic electron microscopy (cryo-EM), where the goal is to reconstruct the 3D potential density of a molecule given multiple noisy tomographic projections obtained at unknown orientations~\cite{mathofcryoem, cryoemreview}. Serving as a simplified yet representative setting, the study of MRA has led to insights about the statistical properties of cryo-EM~\cite{samplecomplexity_cryoem}, and finds practical applications in certain methods, such as class averaging~\cite{classaveraging}.

Importantly, MRA admits a range of informative shift-invariant statistics, sidestepping the problem of shift estimation. Among all possible shift-invariant statistics, three have emerged as the most useful: the \emph{sample mean}, \emph{sample power spectrum}, and \emph{sample bispectrum}. These statistics correspond to the first, second, and third sample moments, and play a central role in method of moments (MOM)-estimators for MRA~\cite{samplecomplexity_mra}.

In this work we adopt a novel Bayesian approach to MRA by learning a diffusion model prior on a class of signals and conditioning this prior on the sample power spectrum. More precisely, we implement the likelihood of the sample power spectrum as a conditioner in a posterior sampling scheme for diffusion models, thereby admitting plug-and-play priors in the form of pretrained models. We refer to the resulting method as moment-based posterior sampling~(MPS). While other orbit recovery methods have included Bayesian components~\cite{EMwithprior}, ours is the first fully Bayesian method.

\section{Related Work} \label{sec:related}

A number of orbit recovery methods for MRA have been studied in the literature, falling into two main categories: MOM-estimators, and (local) maximum likelihood estimators, with some methods combining elements from both~\cite{momentconstraintsphaserecovery}.

In the category of MOM-estimators there are several algorithmic methods that invert the sample moments to recover the orbit~\cite{bispecinv}. While the sample bispectrum is not a sufficient statistic for the orbit of~$X$ in general, it is informative enough to enable consistent estimation of generic~$X,$ where bispectrum inversion methods achieve the optimal sample complexity~$\sigma^6$~\cite{samplecomplexity_mra}. With a sparsity condition on~$X$ it has been shown that the sample power spectrum is sufficient for generic orbit recovery, with a sample complexity of~$\sigma^4$~\cite{sparsesamplecomplexity}.

The category of maximum likelihood estimators is further subdivided into those that estimate the shifts, e.g., angular synchronization~\cite{synchronization}, and those that marginalize the shifts, e.g., expectation--maximization~\cite{EMoriginal}. While most of the methods in the literature are entirely frequentist, a notable exception is expectation--maximization with a prior, representing a local maximum a posteriori estimator~\cite{EMwithprior}.

A common variant of the MRA problem stated in~\eqref{eq:problem_MRA} instead assumes that the distribution of shifts is not uniform, and that this distribution is either known or unknown. In the case of a known, sufficiently non-uniform shift distribution, MOM-estimators can dispense with the sample bispectrum and still recover generic orbits~\cite{nonuniformshifts}. In the case of an unknown shift distribution, Bayesian methods have been applied with explicit priors on the shift~\cite{deeporbitrecKhoo}.

\section{Sample Power Spectrum} \label{sec:powerspectrum}

With~$\mathcal{F}$ as the unitary discrete Fourier transform matrix, the sample power spectrum~$\hat{\mathcal{P}} \in \mathbb{R}^L$ is defined by
\begin{equation*}
    \hat{\mathcal{P}}[\comp] \coloneqq \textstyle \frac{1}{\data} \sum_{\idx = 1}^\data | \mathcal{F} \, Y_\idx [\comp] |^2, \quad \comp \in [\len],
\end{equation*}
where~$[\len]$ denotes the set~$\{0, ..., \len-1\}.$ Note that the shift matrix~$S$ is circulant, so the convolution theorem implies that~$\mathcal{F}$ diagonalizes~$S.$ In particular, for all~$x \in \mathbb{R}^L$ we have
\begin{equation*} \label{eq:shift_dft}
    \mathcal{F} S x [\comp] = \exp{(- 2 \pi i \comp / \len)} \, \mathcal{F} x [\comp] \quad \forall \comp \in [\len],
\end{equation*}
from which it follows that~$\hat{\mathcal{P}}$ is a shift-invariant statistic. However,~$\hat{\mathcal{P}}$ is by itself not sufficient for recovering the orbit of~$X,$ since it carries no information about the phases of~$\mathcal{F} X.$ Nonetheless, an appropriate prior on~$X$ can ensure that the posterior distribution of~$X \mid \hat{\mathcal{P}}$ concentrates near the orbit of the underlying signal, even though posterior estimates may not converge in the data limit. 

To perform Bayesian inference, we require the likelihood of~$\hat{\mathcal{P}} \mid X.$ Simple steps show that~$\hat{\mathcal{P}}[-\comp]=\hat{\mathcal{P}}[\comp]$ for all~$\comp \in [\len]$, with indices understood modulo~$\len.$ The remaining~$\lfloor\len/2+1\rfloor$ components of~$\hat{\mathcal{P}}$ are conditionally independent given $X,$ and are distributed as
\begin{equation} \label{eq:pwrspec_dist}
    \hat{\mathcal{P}}[\comp] \mid X \sim \textstyle \frac{\sigma^2}{\eta_\comp \data} \chi_{\eta_\comp \data}'^{\, 2} \Big{(} \frac{\eta_\comp \data}{\sigma^2} \big{|} \mathcal{F} X [\comp] \big{|}^2 \Big{)},
\end{equation}
where~$c \cdot \chi_{\alpha}'^{\, 2}(\lambda)$ denotes the non-central $\chi^2$-distribution with non-centrality parameter~$\lambda$ and order~$\alpha,$ scaled by~$c > 0,$ while~$\eta_\comp=1$ if~$\comp=0$ or~$\comp=\len/2,$ else~$\eta_\comp=2.$ This likelihood is sufficiently well-behaved to guarantee the existence of a posterior for a wide range of prior distributions~\cite{wellposedbayesinv}.

\section{Moment-based Posterior Sampling} \label{sec:method}

Given the above likelihood for~$\hat{\mathcal{P}} \mid X$ and an appropriate diffusion model prior over~$X$, we can construct a scheme for approximate sampling from the posterior~$X \mid \hat{\mathcal{P}}$.

\subsection{Posterior sampling scheme} \label{sec:method:subsec:aps}

To obtain conditional samples from diffusion models, we implement an approximate posterior sampling scheme with 
the likelihood given in~\eqref{eq:pwrspec_dist}. Specifically, we restrict our attention to score-matching Langevin dynamics (SMLD) models and apply controllable generation~\cite{song2021scorebased}. SMLD models can be viewed as approximating the solution to a backward variance-exploding SDE from~$t=1$ to~$t=0,$ given by
\begin{align*} 
    dX_t &= -\frac{d\sigma_t^2}{dt} \nabla_{\!\!x} \ln p_t(X_t) dt + \sqrt{\frac{d\sigma_t^2}{dt}} dW_t, \\ 
    X_1 &\sim p_1(x) \approx \mathcal{N} \hspace{-1mm} \left ( 0, \sigma_1^2 I \right ), \quad X_0 \sim p_X(x),
\end{align*}
for some data distribution~$p_X,$ where~$W_t$ is a Wiener process and~$t \mapsto \sigma_t^2$ is an increasing diffusion schedule. SMLD solves this SDE by replacing the score~$\nabla_{\!\!x} \ln p_t(x)$ with a data-driven score model~$s_\theta(x, t).$ 

Since the true posterior of~$X_0 \mid \hat{\mathcal{P}}$ is intractable, we apply the principles of controllable generation to the conditioned SDE, approximating the probability paths~$p_t(X_t \mid \hat{\mathcal{P}})$ with
\begin{equation*}
    \nabla_{\!\!x} \ln p_t(X_t \mid \hat{\mathcal{P}})
    \approx s_\theta(X_t, t) + \nabla_{\!\!x} \ln p_t \big{(} \hat{\mathcal{P}}_t' \mid X_t \big{)}.
\end{equation*}
Here~$\hat{\mathcal{P}}_t'$ is a Monte Carlo sample of~$\hat{\mathcal{P}}_t \mid \hat{\mathcal{P}},$ with components sampled independently from the approximate distributions
\begin{equation*}
    \hat{\mathcal{P}}_t[\comp] \mid \hat{\mathcal{P}} \,\overset{\text{approx}}{\sim} \textstyle{\frac{\sigma_t^2}{\eta_\comp \data}} \chi_{\eta_\comp \data}'^{\, 2} \big{(} \eta_\comp \data \max \! \big{\{} \hat{\mathcal{P}}[\comp] - \sigma^2, 0 \big{\}} \big{)}, 
\end{equation*}
while $p_t(\hat{\mathcal{P}}_t \mid X_t)$ is approximated by assuming that~$\hat{\mathcal{P}}_t \mid X_t$ has independent components distributed according to
\begin{equation*}
    \hat{\mathcal{P}}_t[\comp] \, \big{|} \, X_t \overset{\text{approx}}{\sim} \textstyle{\frac{\sigma^2 + \sigma_t^2}{\eta_\comp \data}} \chi_{\eta_\comp \data}'^{\, 2} \big{(} \eta_\comp \data | \mathcal{F} X_t [\comp] |^2 \big{)}.
\end{equation*}
Then, solving the conditioned SDE with the Euler--Maruyama method yields approximate samples of~$X \mid \hat{\mathcal{P}}.$ The complete MPS method is summarized in Alg.~\ref{alg:MPS}.

\begin{algorithm}[h] 
\caption{Moment-based posterior sampling}
\begin{algorithmic}
\State \textbf{Input:} score model $s_\theta(x, t),$ diffusion schedule~$\sigma_t^2,$ step size~$\Delta t,$ sample power spectrum $\hat{\mathcal{P}}$
\State \textbf{Initialize:} sample $X_1' \sim \mathcal{N}(0, \sigma_1^2 I)$
\For{$t = 1, 1-\Delta t, \dots, \Delta t$}
    \State sample $\hat{\mathcal{P}}_t' \sim p_t(\hat{\mathcal{P}}_t \,|\, \hat{\mathcal{P}}), \,\, \xi_t \sim \mathcal{N}(0, I)$
    \State $\tilde{s}(X_t', t) \gets s_\theta(X_t', t) + \nabla_x \ln p_t(\hat{\mathcal{P}}_t' \,|\, X_t')$
    \State $\epsilon_t \gets (\sigma_t^2 - \sigma_{t - \Delta t}^2) / \Delta t$
    \State $X_{t-\Delta t}' \gets X_t' + \epsilon_t \tilde{s}(X_t', t) + \sqrt{2 \epsilon_t}\, \xi_t$
\EndFor
\State \textbf{Output:} Posterior sample $X_0'$
\end{algorithmic}
\label{alg:MPS}
\end{algorithm}

\subsection{Remarks} \label{sec:method:subsec:remarks}

One important advantage of MPS is the wide variety of diffusion model priors that can be admitted. While we only consider SMLD models in this work, it is straightforward to extend MPS to DDPM models~\cite{song2021scorebased} as well. Another advantage is that access to independent posterior samples enables a systematic approach to uncertainty quantification, since any posterior quantity of interest, e.g., posterior covariance, can be estimated to arbitrary precision, assuming that sufficiently many samples are obtained.

It should be noted that the full likelihood of the MRA sample -- while more informative -- would be computationally infeasible to implement. Each observation~$Y_n$ is a Gaussian mixture with~$\len$ components, implying that score evaluation requires~$\mathcal{O}(\data\len^2)$ operations, compared to~$\mathcal{O}(\len)$ operations for the sample power spectrum. This improved computational efficiency is another advantage of MPS over other approaches to Bayesian inference for MRA. In addition, note that the likelihoods of other sample moments, such as the sample bispectrum, could be implemented in MPS.

As stated, our posterior sampling scheme is based on a single Monte Carlo sample per iteration and several simplifying assumptions. To ensure that the posterior is accurately represented, it may be preferable to train a conditioning model~$s_{\Psi}^*(X_t, \hat{\mathcal{P}}, t) \approx \nabla_{\!\!x} \ln p_t(\hat{\mathcal{P}} \mid X_t),$ or to adopt a posterior sampling scheme with consistency guarantees~\cite{yazidcondsample}. Nonetheless, the current scheme produces sufficiently accurate posterior samples for the purposes of this work, and MPS can easily be adapted to other posterior sampling schemes.

\section{Experiments} \label{sec:experiments}

\begin{figure*}[t!] 
    \begin{minipage}[t]{0.49\linewidth}
        \centering
        \centerline{\includegraphics[trim=0 10 0 30, clip, width=\linewidth]{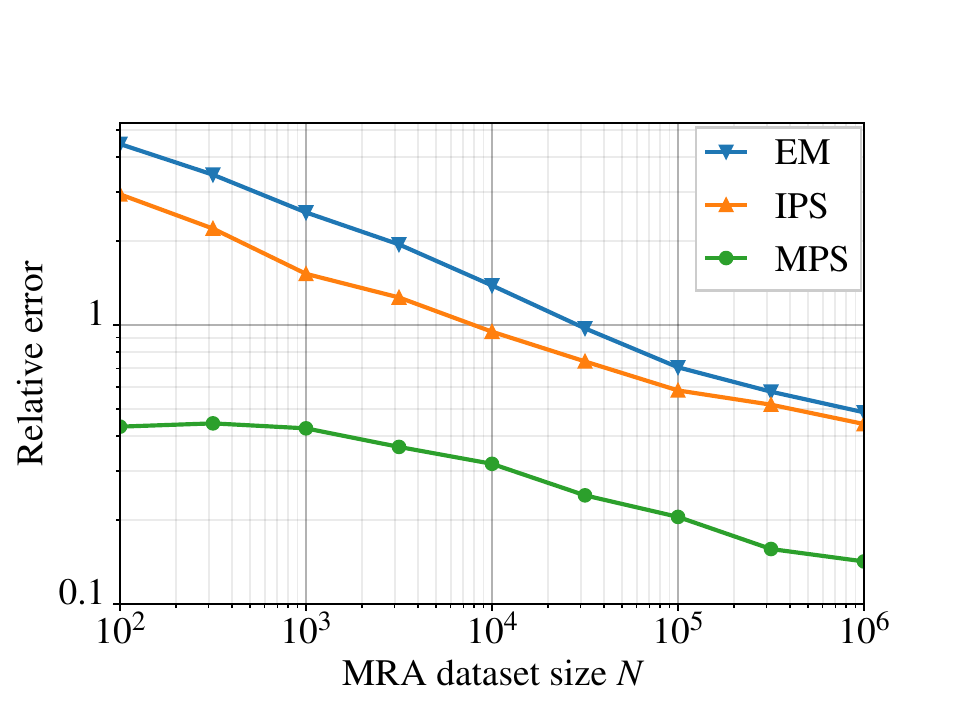}}
        \centerline{\,\,\,\textbf{(a)}}
    \end{minipage}
    \begin{minipage}[t]{0.49\linewidth}
        \centering
        \centerline{\includegraphics[trim=10 30 0 30, clip, width=\linewidth]{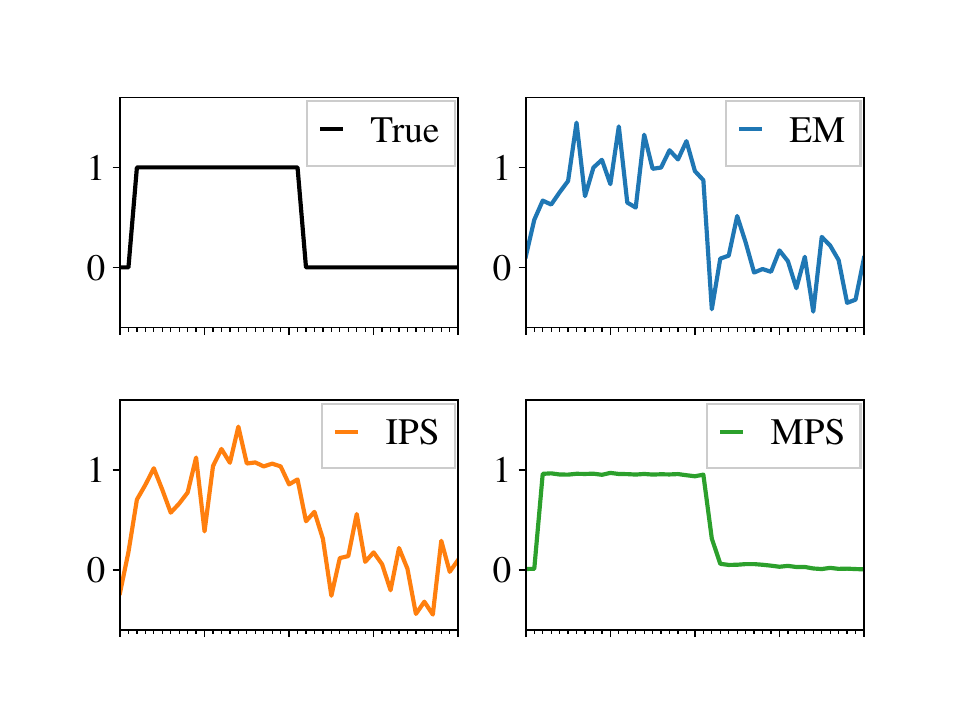}}
        \centerline{\textbf{(b)}}
    \end{minipage}
    \caption{\textbf{(Step signal)} \, \textbf{(a)}: Average relative error of step signal estimates by EM, IPS, and MPS against MRA dataset size~$N.$ \textbf{(b)}: The true step signal~$\truesig$ and aligned estimates by EM, IPS, and MPS for~$N = 10^6.$}
    \label{fig:step}
\end{figure*}

\begin{figure*}[t!] 
    \begin{minipage}[t]{0.49\linewidth}
        \centering
        \centerline{\includegraphics[trim=0 10 0 40, clip, width=\linewidth]{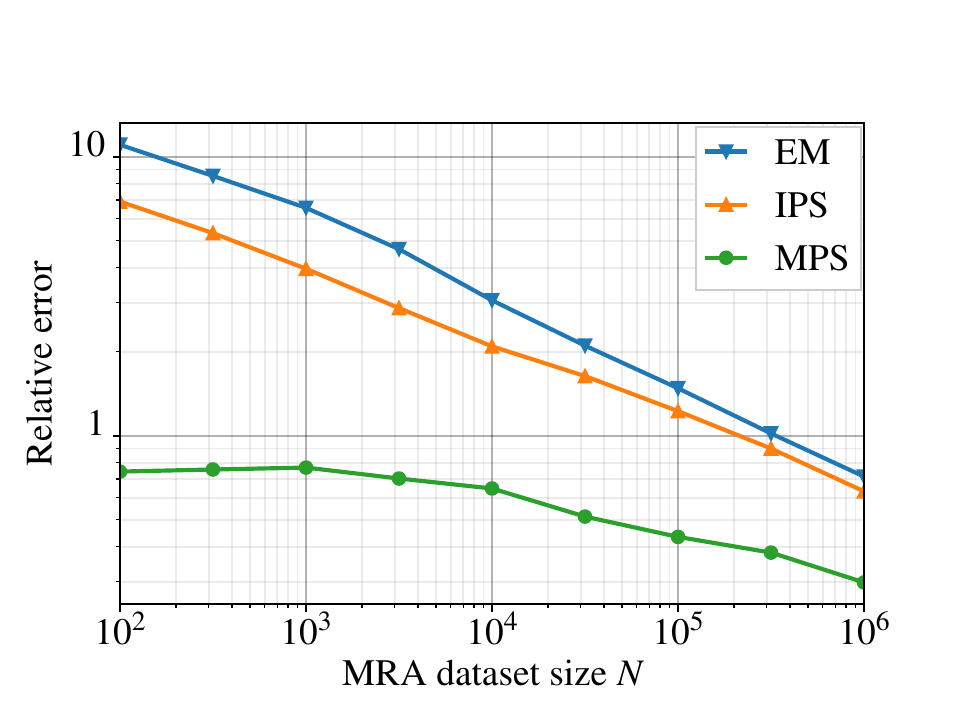}}
        \centerline{\,\,\,\textbf{(a)}}
    \end{minipage}
    \begin{minipage}[t]{0.49\linewidth}
        \centering
        \centerline{\includegraphics[trim=10 30 0 40, clip, width=\linewidth]{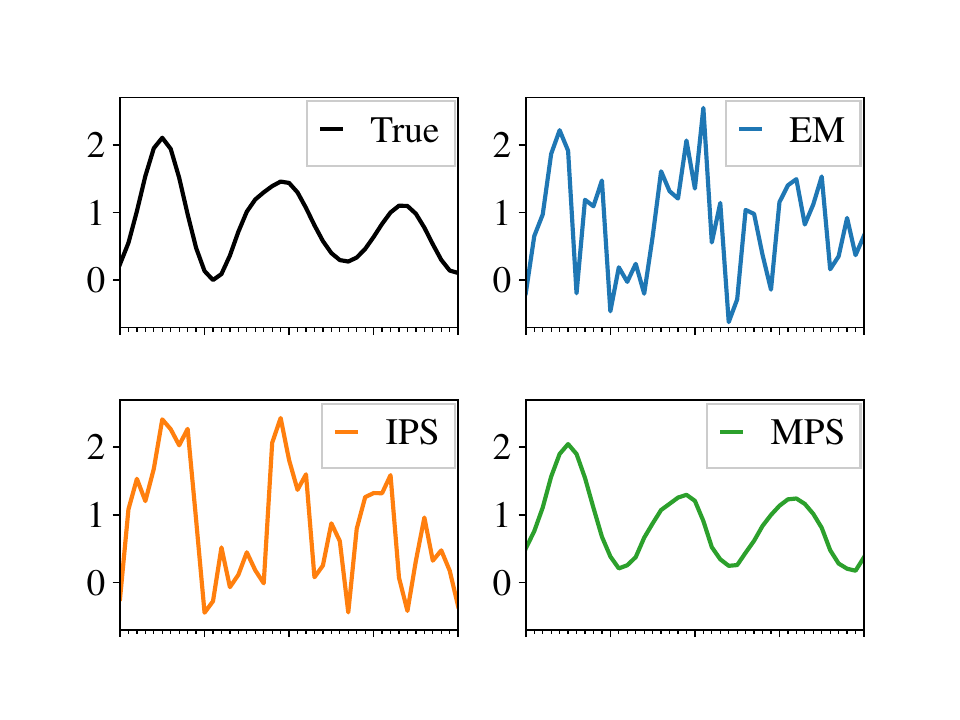}}
        \centerline{\textbf{(b)}}
    \end{minipage}
    \caption{\textbf{(Bell signal)} \, \textbf{(a)}: Average relative error of bell signal estimates by EM, IPS, and MPS against MRA dataset size~$N.$ \textbf{(b)}: The true bell signal~$\truesig$ and aligned estimates by EM, IPS, and MPS for~$N = 10^6.$}
    \label{fig:bell}
\end{figure*}

We validate our approach with numerical experiments that compare MPS to two successful frequentist methods: the first is expectation--maximization (EM), the second is a bispectrum inversion method called iterative phase synchronization (IPS)~\cite{bispecinv}. For the implementations of EM and IPS we follow Bendory et al.~\cite{bispecinv}.

\subsection{Setup and metrics} \label{sec:experiments:subsec:setupmetrics} 

We perform two experiments, distinguished by their respective signal distributions~$p_X.$ The first is termed the \emph{step prior}, and is uniformly distributed on the~$\len$ unit step functions in~$\mathbb{R}^{\len}.$ The second is termed the~\emph{bell prior}, and is a distribution over sums of varying numbers of randomly scaled and translated bell functions in~$\mathbb{R}^{\len},$ with components given by
\begin{equation*}
    X[\comp] \coloneqq \textstyle{\sum}_{k=0}^{K}\, A_k \frac{1}{\sqrt{2 \pi \tau_k^2}} \exp{\left ( - \frac{1}{2 \tau_k^2} | (\comp - C_k) \!\!\!\! \mod{\len} |^2 \right )}, 
\end{equation*}
where~$K \sim \text{Poisson}(10)$,~$A_k \sim \unif([0, 3])$,~$\tau_k^2 \sim \chi_{10}^2$, and~$C_k \sim \unif([0, \len])$, all independent of each other. In both experiments we set~$\len = 41.$ 

Since MPS requires an SMLD model as prior, we train a score model~$s_\theta$ on samples of~$p_X$ until approximate convergence~\cite{song2021scorebased}. For practical reasons the expectation of~$p_X$ is subtracted from training samples, and is therefore added to model samples. In both experiments the model architecture consists of~$8$ shift-equivariant convolution layers with~$4$ channels each, all connected by ReLU activations, resulting in~$10^6$ model parameters. The diffusion schedule is set to~$\sigma_t^2 = 3^{2t}.$ In all other respects the two experiments are equivalent.

We begin by selecting the \emph{true signal}~$\truesig$ with~$\|\truesig\|_2^2 \approx L$ from the support of~$p_X,$ and set~$\sigma^2 = 100$ across all experiments, emulating the worst-case signal-to-noise ratio of cryo-EM data~\cite{cryoemreview}. Next, we generate MRA datasets of increasing sizes~$\data \in \{10^2, \ldots, 10^6\}$ according to~\eqref{eq:problem_MRA} and compute the sample power spectrum~$\hat{\mathcal{P}}.$ We then apply Alg.~\ref{alg:MPS} with the corresponding score model~$s_\theta,$ diffusion schedule~$\sigma_t^2$, step size~$\Delta t = 5 \times 10^{-5}$, and~$\hat{\mathcal{P}},$ generating~$J=2^{10}$ posterior samples~$X_j'$ for every dataset.
The metric of interest is aligned distance, defined as
\begin{equation*}
    d(x, y)  \coloneqq \textstyle \min_{\phi \in [\len]} \| S^{\phi} x - y \|_2,
\end{equation*} 
Hence, we take the MPS estimate~$\hat{X}_{\text{MPS}}$ to be the posterior sample that minimizes internal aligned distance, defined as
\begin{equation*}
    \hat{X}_{\text{MPS}} \coloneqq \textstyle \argmin_{X_k'} \sum_{j=1}^J d \! \left ( X_j', X_k' \right ) .
\end{equation*}

For the final result we report the relative error for each estimate i.e.,~$d(\hat{X}, \truesig)/\| \truesig \|_2$ for~$\hat{X}_{\text{EM}}$,~$\hat{X}_{\text{IPS}}$, and~$\hat{X}_{\text{MPS}}.$ The results of each experiment are averaged over~$20$ repetitions, all with same true signal~$\truesig.$ All experiments were run on an NVIDIA RTX 2000 GPU, with MPS requiring~$\sim\!4$ minutes to generate a set of posterior samples in parallel.

\subsection{Results} \label{sec:experiments:subsec:results}

The experimental results for the step signal and the bell signal are shown in \Fig~\ref{fig:step}\hspace{0.5mm}(a) and~\ref{fig:bell}\hspace{0.5mm}(a), respectively, while \Fig~\ref{fig:step}\hspace{0.5mm}(b) and~\ref{fig:bell}\hspace{0.5mm}(b) display the corresponding true signals along with the best aligned estimates by EM, IPS and MPS for~$N=10^6.$ Both EM and IPS converge at the expected rates as shown by Bendory et al.~\cite{bispecinv}.

Notably, MPS achieves the lowest error across all dataset sizes~$N$ and both signals. For small~$N$ the prior dominates the MPS estimate, and its strength can be deduced from the error at~$N=10^2.$ For large~$N$ the relative error of the MPS estimate decreases significantly, indicating that the sample power spectrum is successfully exploited. Moreover, the best MPS estimate is remarkably similar to the true signal for both priors, and most posterior samples have the same characteristic shape as the true signal (not shown). 

\section{Conclusion} \label{sec:conclusion}

We have introduced MPS, a fully Bayesian method for MRA that obtains posterior samples of the signal by conditioning diffusion model priors on the sample power spectrum. By leveraging pretrained diffusion models, our approach admits a wide array of flexible plug-and-play priors that need not have been trained with this application in mind. Our experiments have demonstrated that MPS can outperform expectation--maximization and bispectrum inversion in the regime of low signal-to-noise ratio and small sample size, highlighting the importance of incorporating prior information in this regime. Taken together, our findings establish a promising framework for other orbit recovery problems, including single-particle cryo-EM. In future work we will explore versions of MPS that exploit more informative sample moments, such as the sample bispectrum.


\vfill\pagebreak


\bibliographystyle{IEEEbib}
\bibliography{strings,refs}

\end{document}